\documentclass[reprint]{revtex4-1}
\usepackage{amssymb}
\usepackage{graphicx}
\usepackage{bm}

\usepackage[latin9]{inputenc}
\usepackage{units}
\usepackage{amsthm}
\usepackage{amsmath}
\usepackage{graphicx}
\usepackage{esint}

\usepackage{color}

\begin{document}

\title{Thermal effects and switching kinetics in silver/manganite memristive systems: Probing oxygen vacancies diffusion}

\author{P. Stoliar}
\affiliation{ECyT, UNSAM, 1650 San Mart\'in, Argentina}
\affiliation{CIC nanoGUNE, 20018 Donostia-San Sebasti\'an, Basque Country, Spain}
\author{M. J. S\'anchez}
\affiliation{Centro At\'omico Bariloche and Instituto Balseiro, CNEA, R\'{\i}o Negro, Argentina}
\affiliation{Consejo Nacional de Investigaciones Cient\'ificas y T\'ecnicas (CONICET), Argentina}
\author{G. A. Patterson}
\affiliation{Instituto Tecnol\'ogico de Buenos Aires (ITBA), Argentina}
\author{P. I. Fierens}
\affiliation{Consejo Nacional de Investigaciones Cient\'ificas y T\'ecnicas (CONICET), Argentina}
\affiliation{Instituto Tecnol\'ogico de Buenos Aires (ITBA), Argentina}

\begin{abstract}
We  investigate  the switching kinetics of oxygen vacancies (Ov) diffusion in  LPCMO-Ag memristive interfaces 
by performing experiments on the  temperature dependence  of the high resistance (HR) state under thermal cycling.
Experimental results are well reproduced  by  numerical simulations based on 
thermally activated Ov  diffusion processes  and  fundamental assumptions  relying on a recent model 
proposed to explain  bipolar resistive switching in manganite- based cells.  
The confident  values obtained  for  activation energies and diffusion coefficient associated 
to Ov dynamics, constitute a validation test  for both   model predictions and   Ov diffusion mechanisms 
in memristive interfaces.
\end{abstract}

\maketitle

Memristive devices (ReRAM) based on oxides compounds are deserving a lot of attention in view of its potential use  
for next generation of non-volatile memories. 
Its operation relies in the resistive switching (RS) effect, which  is the change in  the resistance of the 
device between two different values, the high resistance state (HRS) and the low resistance state (LRS), by an appropriate electric stimulus.
\cite{meijer_2008,sawa} The transition from HRS to LRS is called a  \textbf{set}, while the opposite process 
is defined as \textbf{reset}. 

A large variety of oxides has been explored  for ReRAM applications, ranging from binary  transition metal 
oxides \cite{nio1,zno1,wo1, cuo1,cuo2} to  complex  ones -- manganites and 
perovskites-like. \cite{tsui_2006,szot_2006,janousch_2007,quintero_2007,marlasca_2011,ghenzi_2012,stoliar_2013}

In complex oxides  based devices the emerged consensus points to the  voltage-driven ion migration 
toward/inward the metal-electrode interfaces, as the  relevant  mechanisms controlling bipolar RS, {\it i.e} 
the voltage polarity dependent switching mode.\cite{sawa}
In particular, oxygen vacancies (ions) have been proposed as the active agents participating in the bipolar RS 
effect. \cite{Nian2007,schen_2011,wang_2012,liu_2012,zazpe_2013}

Ref.~\onlinecite{Nian2007} constitutes one of the first evidences supporting  oxygen vacancies (Ov) diffusion in complex oxides 
based devices. 
By electric pulsing Pr$_{0.7}$Ca$_{0.3}$MnO$_3$ films deposited in an oxygen-deficient ambient, and analyzing  the  relaxation in time of the  HR state, information on the  activation energy
 for Ov  diffusion has been obtained.

In addition works searching for ReRAM performance improvement by introducing dopants, have been  also reported  consistently. 
As an example, the efficiency  of Ov migration by Nb doping Ba$_{0.7}$Sr$_{0.3}$TiO$_3$ (BST) thin films, 
was tested in Ref.~\onlinecite{jung_2012}. 
As stated in  that work, the  defects distribution is strongly related to the RS properties, assisting the Ov 
migration and making more efficient the ReRAM operation. 

Although  the exact microscopic origin  behind the  RS effect remains elusive, a recent phenomenological model, 
named voltage enhanced oxygen vacancy (VEOV) migration model, \cite{rozen_2010} succeeded in reproducing many non trivial characteristic of bipolar RS experiments carried in complex oxides,
under different stimulus protocols. \cite{ghenzi_2010,marlasca_2011,ghenzi_2012} 
The VEOV model incorporates as main ingredients (i) the drift/diffusion of Ov along the highly resistive metal/oxide interfaces, 
 where strong electric fields developed and (ii) a linear relation between resistivity   and Ov concentration.
 Thus, according to the model, the local change in the concentration of  Ov near the electrodes modifies the contact  resistance, 
as follows from (ii). 
Depending on the polarity of the electric field during a switching operation, Ov  might  accumulate  or 
void nanosized  regions at the metal/electrodes interfaces, giving place to the  reset and set  transition respectively.

In spite of the success of the VEOV  model in reproducing many experimental features of bipolar resistive switching in complex oxides, the drift/diffusion 
of Ov  has  been only tested indirectly.\cite{marlasca_2011,stoliar_2013}

The goal of the present  work is  to probe the switching kinetics proposed in the  VEOV model, by studying the response of the switched  HR state  under thermal cycling  in  a  manganite (LPCMO)-Ag  memristive  interface. 
As we describe below the  experiments  are contrasted with  numerical simulations based on the  VEOV model 
\cite{rozen_2010} from which we extract confident values for  the activation energies and diffusion coefficient associated 
to Ov dynamics. 
In this way a new validation test, for  both  model predictions and thermal activated Ov diffusion mechanisms, 
is performed.

We conducted experiments on two manganites $\mathrm{La_{5/8-y}Pr_{y}Ca_{3/8}MnO_{3}}$ (LPCMO)-based samples. 
This compound exhibits bipolar RS when an external stimuli is applied.\cite{Levy2002} 
Contacts were made by depositing drops of silver paint of 1--2 mm diameter over the LPCMO pellet. 
Each sample was coupled to a different heating stage in order to perform two  completely independent analyses 
in different laboratories.
In setup \#1 temperature was controlled by a LakeShore 331 Temperature Controller, having an
error lower than 1 K. Setup \#2 consisted of an \textit{ad hoc} heating stage based on a Peltier cell 
controlled by an Arduino board.

\begin{figure}
\includegraphics[width=\columnwidth]{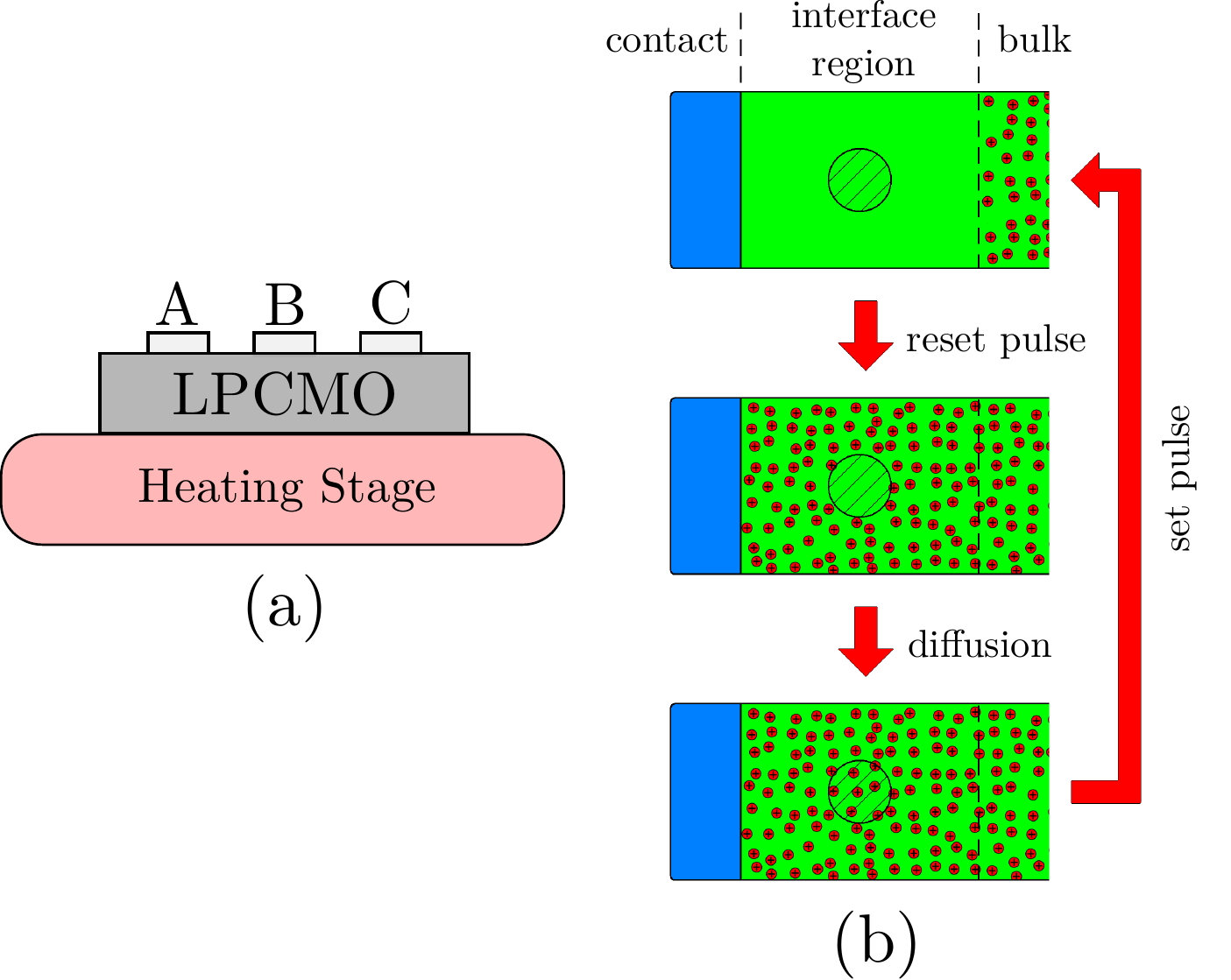}
\caption{\label{fig:Experimental-setup} (a)  Scheme of the experimental setup. 
(b) Model description and proposed mechanism described in the text. 
While the sample is in the LR state,  Ov are  mostly in the bulk. 
Under  a reset pulse, the Ov move to the interface region and the HR state is attained. 
When the pulse is switch off, the temperature activated diffusion process can increase even more the interface resistance.}
\end{figure}

In both experiments the HR and LR states of a single interface -- the one corresponding to the contact A 
in Fig.~\ref{fig:Experimental-setup}(a) -- were measured. 
However in order to extract information about the Ov switching dynamics,  
we shall solely analyze   the temperature dependence of the HR state, $R_{A}$.  
-this election will be  justified below, when we describe the 
Ov dynamics at the interface.

The general procedure employed  for the experiments consists of applying a writing  reset pulse at an 
initial temperature and then recording  the HR state, $R_{A}\left(T\right)$, during several cyclic temperature sweeps. 
For each temperature, the resistance was measured  applying a small bias current between terminals A and C and 
recording the voltage drop between A and B.

\begin{figure}
\includegraphics[width=\columnwidth]{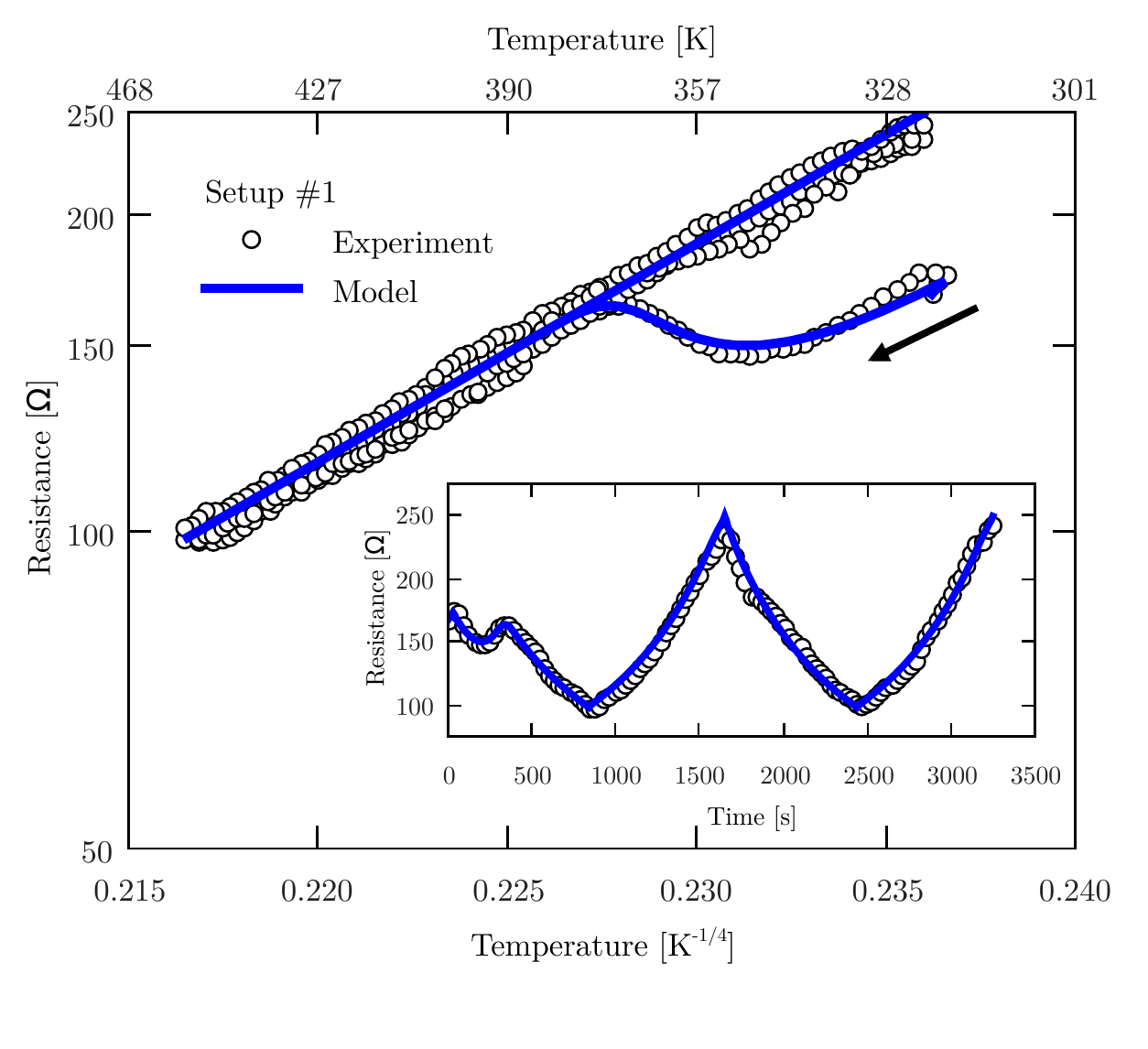}
\caption{\label{fig:fit_data2}(Color online) Experimental results for setup \#1 showing the temperature behavior of the contact  HR,  $R_{A}(T)$. 
The arrow indicates the starting direction. 
At $T\gtrsim 373$ K the sample is fully diffused. 
The solid blue line is the model fit. 
Inset: time evolution of $R_{A}(T)$ and model fit.}
\end{figure}

In Fig. \ref{fig:fit_data2} we present results corresponding to setup \#1. 
We swept the temperature from 323 K to 455 K with a rate of $\pm$10 K/min. 
Results are plotted in semi-log scale \textit{vs}. $T^{-\nicefrac{1}{4}}$. 
Indeed, throughout this work we use the variable range-hopping model to account for the semiconductor-like behavior 
of the samples, 

\begin{equation}
R=R_{0}\exp\left(\frac{T_{0}}{T}\right)^{\nicefrac{1}{4}},
\label{eq:sigma}
\end{equation}
\noindent
where the characteristic temperature $T_{0}$ depends on the localization of the charge carriers and $R_{0}$ is a scaling 
factor. \cite{Viret 1997,Ambegaokar1971} 
It is true that the temperature range of our experiments is far is too small to rule out other transport mechanisms; 
yet, we find that it matches slightly better to our data, and it has been proposed for manganites in the paramagnetic 
region. \cite{Viret 1997,Sun2000} The inset of Fig.~\ref{fig:fit_data2} shows the time evolution of $R_{A}$ 
associated to the temperature sweep.

As it can be seen from  Fig. \ref{fig:fit_data2}, during the first heating ramp, the HR state follows 
an unexpected path eventually arriving to a stable state that matches the VRH behavior  Eq. (\ref{eq:sigma}).
This stable HR state is reached at $T \gtrsim $ 373 K.

\begin{figure}
\includegraphics[width=\columnwidth]{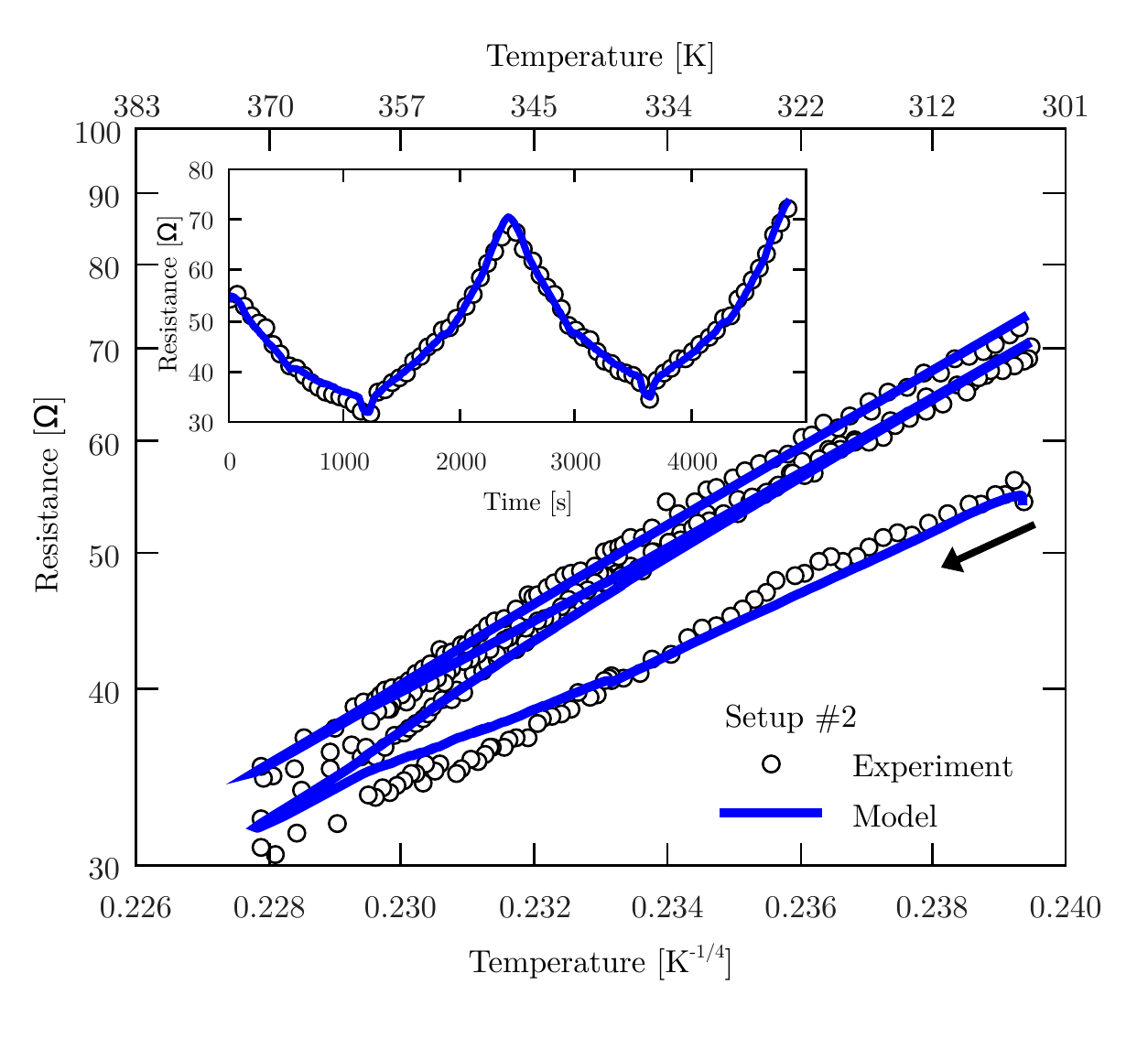}
\caption{\label{fig:fit_data3}(Color online) Experimental results for setup \#2 showing the temperature behavior of the contact resistance $R_{A}(T)$. 
The arrow indicates the starting direction. 
The sample continues the diffusion process for more than one cycle. 
The solid blue line is the model fit. Inset: time evolution of $R_{A}(T)$ and model fit.}
\end{figure}

We repeated the experiment with setup \#2, this time with temperature in a narrower range, 
from 303 K to 373 K with a rate of up to $\pm$6 K/min.
Results are presented in Fig.~\ref{fig:fit_data3}. In principle, a qualitative similar behavior is obtained: 
below $T \approx $ 373 K, the resistance deviates from the VRH prediction.
However, in contrast to the data in Fig.~\ref{fig:fit_data2}, after the first heating ramp, the HR did not completely match 
the VRH prediction.

The above results suggest that there is a temperature activated diffusion  that changes the local resistivity in the interface.

\begin{figure}
\includegraphics[width=\columnwidth]{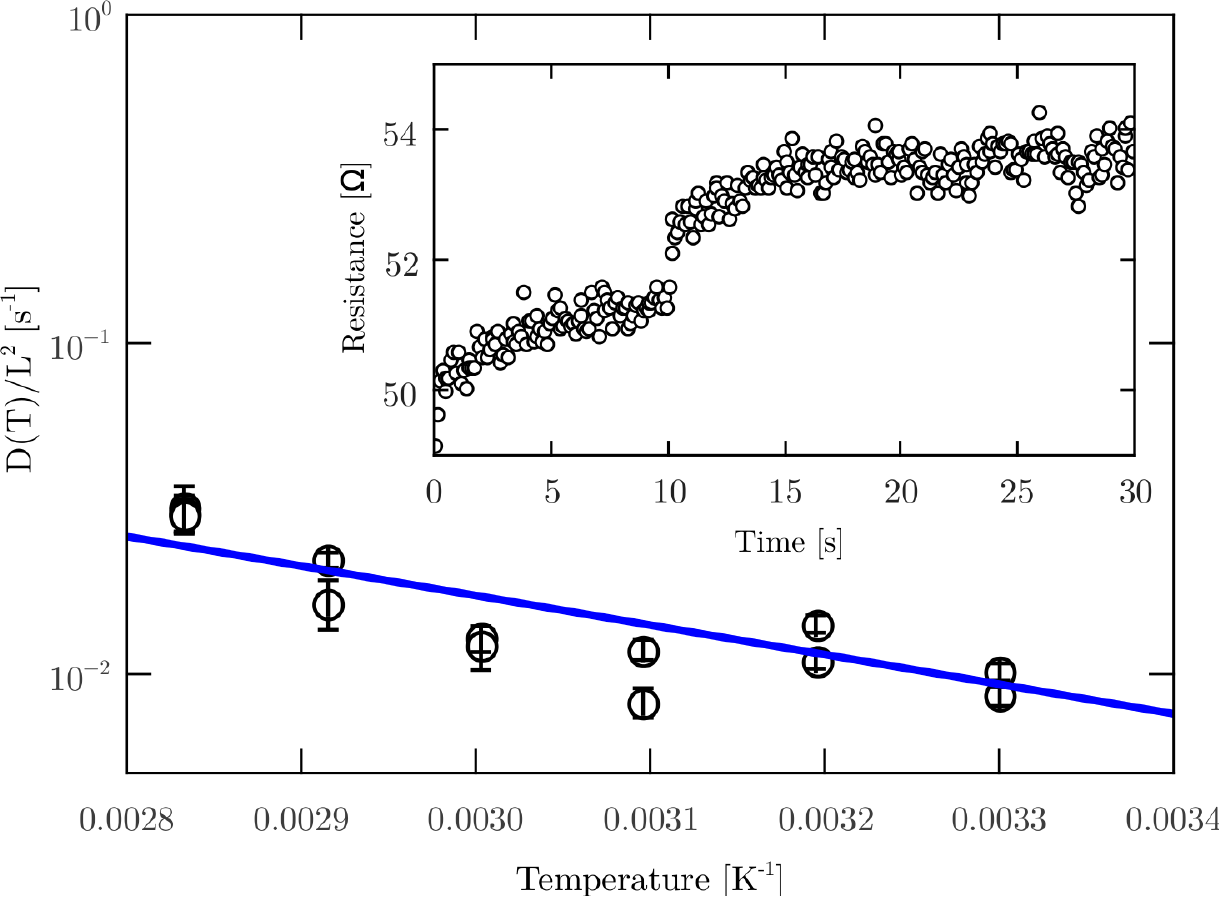}
\caption{\label{fig:fit_ea}(Color online) Diffusion coefficient as a function of temperature. 
The activation energy obtained is $(0.18 \pm 0.03)$ eV. Inset: time evolution of the HR state at 353 K. 
Two consecutive diffusion processes are observed at this temperature.}
\end{figure}

Based on the VEOV model predictions, our interpretation of the obtained behavior is the following. 
The negative polarity of the reset pulse forces Ov to move from the bulk to the interface region, increasing 
its resistance $R_A$.\cite{marlasca_2011} 
However, the Ov distribution is far from equilibrium; 
the reset pulse does not succeed in introducing vacancies homogeneously in all the interface and some 
nanoscale regions remain void of vacancies. After the reset pulse, the dynamics is governed by diffusion.
However at room temperature, the diffusion coefficient $D(T)$ for Ov is too small 
to effectively compensate the density gradient obtained as a consequence of the reset pulse. 
As temperature is raised, $D(T)$ rapidly increases enabling former void regions to be infilled with Ov,
further producing the concomitant increase in the resistance. 
This process lasts until the steady configuration is attained. 
A picture of the proposed mechanism is  shown in Fig.\ \ref{fig:Experimental-setup}(b).

To be definite we model a nanoscale regions at the interface with defect of vacancies in one dimension, as 
a region delimited between $- L<x<L$ with an uniform Ov density profile $n$= $n_b$ everywhere but 
in a void region $(-x_o,+x_o)$, 
%
\begin{equation}
n\left(x,t=0\right)=\begin{cases}
0 & x\in (-x_o,+x_o),\\
n_{b} & x\in (-L,-x_o]\cup[+x_o,+L),\end{cases}
\label{eq:initial}
\end{equation}
where $x$ is the 1-D spatial coordinate and $t$ is the time.

We do not expect the results to be qualitatively different for a different choice of $n\left(x,t=0\right)$, 
as long as it contains a strong discontinuity. 
In addition the extension to 2D and 3D is straightforward in cartesian coordinates. 
Vacancy infilling dynamics is governed by Fick's second law,
\begin{equation}
\frac{\partial n}{\partial t}=D\left(T\right) \frac{\partial^{2}n}{\partial x^{2}}\qquad x\in (-L,+L),\;t>0.
\label{eq:fick}
\end{equation}
The diffusion coefficient $D(T)$ is thermally activated with activation energy $\epsilon$
\begin{equation}
D\left(T\right)=D_{0}\exp\left(\frac{-\epsilon}{k_{B}T}\right),
\label{eq:d}
\end{equation}
where the prefactor $D_{0}$ is the diffusion coefficient at infinite $T$ and $k_{B}$ the Boltzmann constant. 

We assume that the reset procedure left the interface with  a slightly different $R_{0}\equiv R_{0,H}^{\prime}$ 
respect to the final HR state (see  Eq. (\ref{eq:sigma})).
The small change from $R_{0,H}^{\prime}$ to $R_{0,H}$ is due to the infill of vacancies.
 Following the VEOV model assumption, we then consider a linear dependence 
between the value of $R_{0}$ and the total number of vacancies 
in the former void region, $N_{i}=\int_{-x_o}^{+x_o} n(x,t)dx$,
\begin{equation}       
R_{0}\approx R_{0,H}^{\prime} + B\ N_{i} \; , 
\label{eq:a+bN}
\end{equation}
where $R_{0,H}^{\prime} $ and $B$ are two arbitrary values. 

Solving Eqs.(\ref{eq:initial})-(\ref{eq:d}) and using Eqs.(\ref{eq:sigma}) and (\ref{eq:a+bN}), we obtain that
\begin{equation}
R\left(t\right)\approx\left[R_{0,H}^{\prime} + B\left(\frac{x_o}{L}-\sum_{k=1}^{\infty} \phi_k\psi(t)\right)\right]\exp\left(\frac{T_{0}}{T}\right)^{\nicefrac{1}{4}},
\label{eq:fitting}
\end{equation}
where 
\begin{equation}
\phi_k=\left(\frac{\sqrt{2}\sin\left(k\pi\frac{x_o}{L}\right)}{k\pi}\right)^2,
\end{equation}
\begin{equation}
\psi(t) = \exp\left\{-\left(k\pi\right)^2\frac{D_0}{L^2}\int_{0}^{t} \exp\left(\frac{-\epsilon}{k_{B}T(t')}\right)dt'\right\}.
\label{eq:timeevol}
\end{equation}

We fit Eq.\ (\ref{eq:fitting}) to the experimental measurements of $R_{A}$ (see Figs.\ \ref{fig:fit_data2}--\ref{fig:fit_data3}). 
First, $T_{0}^{\nicefrac{1}{4}}$ was  obtained from stable regions in the experiments presented 
in Figs. \ref{fig:fit_data2} and \ref{fig:fit_data3}, \textit{i.e.}, long after transients have passed.
Fitted values of $T_{0}^{\nicefrac{1}{4}}$ were in the range of $[45, 57]\; \mathrm{K}^{\nicefrac{1}{4}}$.
Using Eq.(\ref{eq:fitting}), we then looked for the values of $\nicefrac{x_o}{L}$, $\nicefrac{D_0}{L^2}$ 
and $\epsilon$ which minimized the mean square error between the experimental results and simulated values.\cite{nr} 

Among different experiments, we obtained $\nicefrac{x_o}{L}\sim 0.5$.
In the case of  $\epsilon$- the height of the barrier that maintains the system in a metastable out-of-equilibrium 
condition immediately after the writing pulse- we got  values ranging between 0.3 -- 1.3 eV. 
The equilibrium cannot be achieved at room temperature; indeed, 
it is necessary to increase the thermal energy by $\sim$25\% in order to significantly activate the displacement of Ov.
Regarding the dispersion in the values, we hypothesize that it is evidencing a broad distribution of anchoring energies instead of the mono-energetic level considered in our model. 
Moreover, experimentally we have little control (if any) in the way this broad distribution of states is infilled.   

Extracting $\nicefrac{D_0}{L^2}$ is not straightforward because it is exponentially affected by the estimated value of $\epsilon$. 
In fact, there is a strong nonlinear covariance between $\epsilon$ and $\nicefrac{D_0}{L^2}$ 
in the proposed minimization method.

Therefore, we estimated $\nicefrac{D(T)}{L^2}$ from the measured time evolution of resistance for a fixed temperature and 
the first term in Eq.(\ref{eq:fitting}).
In the inset of Fig. \ref{fig:fit_ea}, we present 
an example of the evolution of the resistance after setting the HR state at 353 K. 
We hypothesize that the two successive exponential-like-increase intervals could correspond 
to the sequential infilling of two vacancy-void regions. Indeed, we also 
obtained good fits (not shown here) of Figs.~\ref{fig:fit_data2}-\ref{fig:fit_data3} with a two-vacancy-void-region model.

Fig.\ \ref{fig:fit_ea} displays the  results for six temperatures (two sets of measurements for each temperature)
together with the fit from which we extracted
$\epsilon \sim 0.18$ eV,  $\nicefrac{D_0}{L^2} \sim 90$ s$^{-1}$.

A value of ${D_0}\sim 2 \cdot 10^{-7}$ $\mathrm{cm^{2}/s}$ is obtained by setting $L = 500$ nm, which is  
fully consistent with reported oxygen diffusion constant in perovskite- based oxides.\cite{Nian2007,watterud} 
In the case of  $\epsilon$ and $T_0$ the fitted values are of the  order but slightly smaller than those
 reported in literature.\cite{Viret 1997} 

In summary, by performing temperature sweeps of the stable HR states in a perovskite- based memristive interface, and relying 
on simple assumptions of the VEOV model, we have extracted confident values of relevant parameters involved in the kinetics 
of Ov diffusion. 
We have found that the electric pulse might set the system in a metastable configuration that relax after overcoming a 
barrier of 0.3 -- 1.3 eV.
We have also extracted diffusion coefficient values for oxygen vacancies that are consistent with literature, 
supporting that they are the responsible for the resistance modulation at the interface. 
Moreover, these findings support a switching dynamics completely in-line with the VEOV model. 

 Finally, the metastability here reported could be also further
exploited in resistive switching binary memories, as it actually increases the resistance of 
the HR state (\textit{i.e.}, it improves the ON/OFF ratio).

This work was partially supported by CONICET (PIP 11220080101821), ANPCyT (PICT-2010 \# 121), Fundaci\'on Balseiro and ITBA (ITBACyT-2013 \# 6).
We acknowledge P.Levy and G.Leyva for providing the LPCMO samples, and for the use of facilities at GIA-CAC-CNEA, Argentina. PS acknowledges the Ram\'on y Cajal program (RYC-2012-01031). GAP and PIF acknowledge F. Sangiuliano Jimka for lab assistance.
The authors would like to thank M. Rozenberg, F. Gomez-Marlasca, N. Ghenzi and C. Acha for useful discussions.


\end{document}